\documentclass[10pt, conference, compsocconf]{IEEEtran}
\usepackage{amsmath, amssymb}
\usepackage{algorithm}
\usepackage{algorithmic}
\usepackage{graphicx}

\begin{document}
\title{Delay Modelling for a Single-hop Wireless Mesh Network under Light Aggregate Traffic}

\author{
\IEEEauthorblockN{Albert Sunny\IEEEauthorrefmark{1}, Joy Kuri\IEEEauthorrefmark{2} and Saurabh Aggarwal\IEEEauthorrefmark{3}}
\IEEEauthorblockA{Center for Electronics Design and Technology\\
Indian Institute of Science, Bangalore-560012, India\\
Email: \IEEEauthorrefmark{1}salbert@cedt.iisc.ernet.in, \IEEEauthorrefmark{2}kuri@cedt.iisc.ernet.in, \IEEEauthorrefmark{3}saggarwal@cedt.iisc.ernet.in}
}

\bibliographystyle{IEEEtran}

\maketitle

\begin{abstract}
In this paper, we consider the problem of modelling the average delay in an IEEE 802.11 DCF wireless mesh network with a single root node under light traffic. We derive expression for mean delay for a co-located wireless mesh network, when packet generation is homogeneous Poisson process with rate $\lambda$. We also show how our analysis can be extended for non-homogeneous Poisson packet generation. We model mean delay by decoupling queues into independent M/M/1 queues. Extensive simulations are conducted to verify the analytical results.	
\end{abstract}


\IEEEpeerreviewmaketitle

\section{\large{Introduction and Related Work}}
\label{intro}

\begin{figure}[h]
\centering
\includegraphics[scale=0.5]{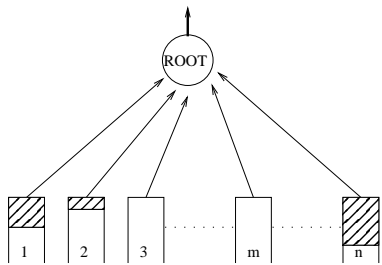}
\caption {A single-hop  wireless mesh network}
\label{fig:tree}
\end{figure}

The IEEE 802.11 has become ubiquitous and gained widespread popularity as a protocol for wireless networks. As a result, various models have been proposed to analyze and model the parameters of interest.

Since the seminal paper by Bianchi \cite{bianchi}, throughput analysis of IEEE 802.11 DCF has come under much scrutiny. In \cite{bianchi}, the author evaluates the aggregate system throughput as a function of the number of nodes under saturation, i.e., each user has a packet to transmit at all times. The main feature of the analysis is the 2-dimensional Markov model, which captures the back-off phenomenon of IEEE 802.11, given an attempt rate for each node. Due to the robustness and simplicity of the model, it has been used extensively by several researchers in extending their work.  In \cite{tay}, the authors present an analytical model for throughput analysis of DCF using
average back-off state as as compared to the Markovian model by Bianchi. Although the approaches are different, the end numerical results are close to each other. In \cite{akumar}, the authors have studied the fixed point solution and performance measure in a more generalized framework.

Delay analysis of IEEE 802.11 DCF is limited in comparison to the throughput studies. In \cite{mmc}, the authors present delay analysis of HOL packet for the saturation case. In \cite{tickoo} and \cite{sikdar}, the authors have extended the model of Tay and Chua \cite{tay} by proposing G/G/1 queues for each individual user. However, the analysis ignores the random delays due to packet transmissions by other users. The authors arrive at an expression for unsaturated collision probability using fixed point analysis and use the same in their subsequent modelling. 

In \cite{tobagi}, the authors have proposed System Centric and User Centric Queuing Models for IEEE 802.11 based Wireless LANs. \cite{tobagi} assumes the server to allocate its resources to users in a round robin manner. In the System Centric Model, the arrivals are assumed to be Poisson, thus the resource sharing model takes the form of an M/G/1/PS system with the mean delay being the same as that in an equivalent M/M/1 system. In the User Centric  Model, each user queue is modeled as a separate G/G/1 queue. 

A novel model based on Diffusion approximations has been used to model delay in  ad-hoc networks  by Bisnik and Abouzeid \cite{bisnik}. The authors, provide scaling laws for delay under probabilistic routing. The authors assume the ad hoc network to follow probabilistic routing methodology (i.e the routing is oblivious to the origin and nature of the packet). In the paper, the authors have considered the problem of characterizing average delay over various network deployments. But for a given network deployment the value of the observed average delay may vary widely in comparison to the value as calculated using the diffusion approximation model. We are interested in a simple model to obtain average delay for a given mesh network as against the average over many random deployments \cite{bisnik}.

In this paper, we derive a simple closed form expression to model the average user traffic delay for single hop wireless mesh networks, 
using the M/M/1 queuing model for homogeneous Poisson packet generation. We remark the user traffic delay is not merely the Head-Of-Line (HOL) packet delay that has been analyzed in \cite{bianchi}, \cite{tay} and \cite{akumar}; it includes the the delay from the time a user packet arrives at the queue, till the packet leaves the node. Thus, both queuing delay and HOL delay are included. The work closest to ours appears in \cite{tobagi}. In \cite{tobagi}, the authors provide a relationship between attempt probabilities and collision probabilities for an \emph{unsaturated} network. Our objective on the other hand, is to explore the use of known results for the \emph{saturated} network in analyzing the mean delay experienced by traffic in a lightly loaded WLAN. Thus,  \cite{tobagi} analyzes the problem from a very basic viewpoint, while our goal is to exploit available results for a ``quick and dirty" assessment of mean delays. In our approach, all the contending queues are assumed to be independent and have been decoupled from one another under the assumption of equal throughput sharing among the contending nodes. Each of the decoupled queues is thereafter modelled as an independent M/M/1 queue for computation of average delay. The main results of this paper can be summarized as follows:

\begin{itemize}
\item We derive a simple expression for average delay for single-hop mesh networks with homogeneous Poisson packet generation.
\item We also extend our results for the case of non-homogeneous Poisson packet generation
\item We show though simulation that our model is valid under low to moderate aggregate arrival rates with reasonable degree of accuracy.
\end{itemize}

We find that even though our analysis makes apparently drastic assumptions, the results match simulations very well when the load on the network is small. For example, in the light to moderate load regime, the error between analysis and simulation is not more than 10\%. Thus, for low to moderate loads, we have a simple analytical formula that captures mean delay with reasonable accuracy. The rest of the paper is organized as follows. In Section II, the system model is described in detail. In Section III, we present the mean delay analysis. In Section IV, the proposed model is validated against simulation results. Finally, Section V presents conclusion and remarks regarding future work.

\section{\large{System model}}
\label{model}
\begin{figure}[h]
\centering
\includegraphics[scale=0.4]{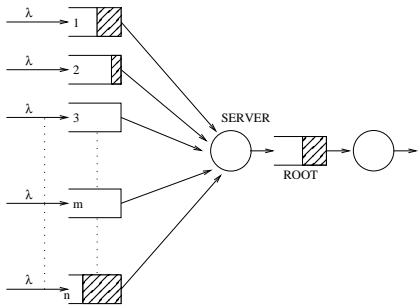}
\caption {System model}
\label{fig:model}
\end{figure}

We consider a single hop mesh network with $n$ nodes and no hidden nodes and a \textit{ROOT} node. We assume the each node in the network generates data an average rate $\lambda$; the destination is \textit{ROOT} node. Moreover, to simplify the analysis, we assume the data generation process to be Poisson with rate parameter $\lambda$. Each node in the network shares the medium and uses the IEEE 802.11 DCF to send data to the root node. We assume the that there is data transfer only on the uplink (i.e from the nodes to the root) on an error free channel. 

From \cite{bianchi}, \cite{tay} and \cite{akumar}, it is known that the average aggregate rate of data transfer is dependent on the number of nodes contending. We also note that each node has equal probability of success. As in \cite{tobagi}, we model the network as a multiple queue single shared server system, where the service rate of the server is dependent on the number of non-empty queues and the server selects a non-empty queue uniformly at random. We assume that the root node can push data out of the network instantaneously. In this system, we are interested in the quantifying the average delay between a packet's arrival to a queue and its departure from the system. 

\section{\large{Delay Modelling}}
\label{analysis}

\subsection{Computing aggregate service rate in 802.11 DCF}

We start this section by discussing the mechanism of 802.11 DCF. In 802.11 DCF, at the end of every successful transmission, a Short Inter frame space (SIFS) is provided for sensing the channel. At the end of SIFS, each of the nodes starts a DCF inter-frame space timer (DIFS). When the DIFS timer expires, each node enters a back-off phase. To prevent collisions, random back-off is used to order the transmissions. Each of the nodes freezes its back off timer in case any one of the nodes starts transmission. Upon completion of transmission, the remainder of the back off is continued by the other nodes. If two or more nodes finish their back-offs within one slot of each other, a collision occurs. On detection of collision, the colliding  nodes sample back-offs from a doubled collision window. A successful transmission occurs if and only if exactly one of the nodes finishes its back-off in a given slot. Soon after, the node starts sending the packet. Upon sensing activity on the channel, other nodes freeze their back-off timers and defer their transmissions until the channel is sensed to be idle again.

Out of the $n+1$ nodes, one node is a \textit{ROOT} and we assume that there is no downlink traffic (i.e from the root to the nodes). Thus, atmost $n$ nodes are contending for access to the wireless medium. Let $\beta_n$ be the probability that a node attempts transmission, when $n$ nodes are contending for access to the wireless medium. From the paper by Bianchi \cite{bianchi}, $\beta_n$ can be expressed in terms of the conditional collision probability $p$ in two ways as follows.

\begin{eqnarray}
\beta_n(p) &=& \frac{2 \cdot (1-2p)}{(W+1) \cdot (1-2p) + pW \cdot (1 - {(2p)}^m)} \label{eq:beta1} \\
\beta_n(p) &=& 1 - (1-p)^{\frac{1}{n-1}} \label{eq:beta2}
\end{eqnarray}

Now, the RHS of Equation \eqref{eq:beta1} is monotonically decreasing from $\frac{2}{W+1}$ to $\frac{2}{2^mW+1}$, for $p \in [0,1]$, and the RHS of Equation \eqref{eq:beta2} is monotonically increasing from $0$ to $1$, for $p \in [0,1]$. We can use fixed point analysis to obtain $\beta$ for a given $n$. Let $T_S, T_I$ and $T_C$ be the durations of success, idle and collision slots, respectively. Let us define the probability of a successful transmission ($p^{(n)}_S$), collision ($p^{(n)}_C$) and idle ($p^{(n)}_I$) as
$$ p^{(n)}_S = n\beta_n \cdot (1-\beta_n)^{n-1}$$
$$ p^{(n)}_I = (1-\beta_n)^n $$
$$ p^{(n)}_C = 1-  p^{(n)}_S - p^{(n)}_I$$	
By applying the Renewal Reward theorem, we define the average throughput in terms of packets per second as follows
$$S(n) = \frac{p^{(n)}_S}{p^{(n)}_IT_I + p^{(n)}_ST_S + p^{(n)}_CT_C}$$
where

\begin{figure}[h]
\centering
\includegraphics[scale=0.3,angle=-90]{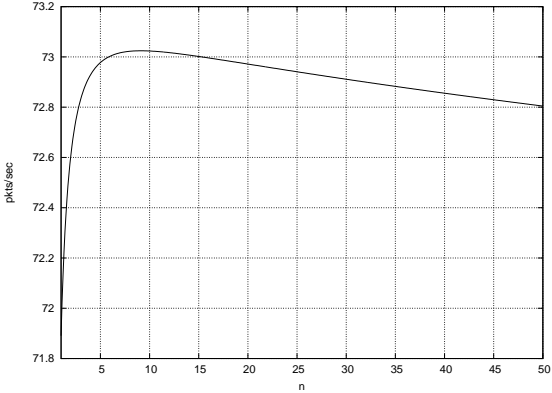}
\caption {Variation of throughput with the number of contending nodes. For the above plot, packets size was set as $1500\,Bytes$ and a transfer rate of $1\,Mbps$ and the parameters were set to mimic 802.11b}
\label{fig:thruput}
\end{figure}

From Figure \ref{fig:thruput}, it can be see that the throughput remains fairly constant over a wide range of $n$.

\subsection{Decoupling of queues}
In general, the service rate per queue may depend on the arrival rate and the number of contending nodes. Now let us assume a constant average throughput $C$ (5 or more nodes in Fig \ref{fig:thruput}). This assumption is justified from the argument in the previous section.  When $N$ users access the medium with different channel data rates, each user achieves $\frac{1}{N}$ of the total long term throughput, regardless of differences in their channel data rate \cite{hue}, \cite{sad}. Now, since each queue has equal probability of succeeding and has the same arrival rate $\lambda$, we can say that the each queue will see an instantaneous average service rate of $\frac{C}{N_s}$ at time $s$, where $N_s$ is the number of non-empty queues at time $s$. So we can see that given that $N_s$ nodes are contending, the instantaneous average service rate for a queue can be modelled as $\frac{C}{N_s}$. Thus our modelling approach is to decouple the queues into $N_s$ queues, each with an arrival rate $\lambda$ and average service rate of $\frac{C}{N_s}$.

\begin{figure}[h]
\centering
\includegraphics[scale=0.5]{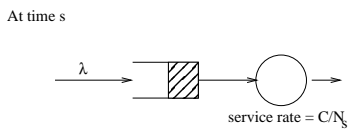}
\caption {A decoupled queue}
\label{fig:decouple}
\end{figure}

It can be shown that the number of slots between two consecutive successes follows a geometric distribution when the system is modeled as a discrete time system. Therefore, in our continuous time model, we take the service times to be exponentially distributed. So, we model the decoupled queues at time $s$ as independent M/M/1 queues each with an arrival rate $\lambda$ and instantaneous service rate of $\frac{C}{N_s}$.

\subsection{Modelling average delay for homogeneous Poisson packet generation}
Let $N_s$ be the number of non-empty queues in the system at time $s$. Then from the previous section, we can argue that the instantaneous  service rate seen by a non-empty queue at time $s$ is given by
$$M_s = \begin{cases} \frac{C}{N_s} &	N_s \neq 0 \\  0 & N_s = 0 \end{cases}$$
We are interested in finding the average service rate over time instances , when $N_s$ is non-zero. In order to do that, we first define
$$\Sigma_t = \{s: 0 \leq s \leq  t, N_s \neq 0\}$$
Let
$$ M_{avg} = \lim_{t \to \infty} \int_{s \in \Sigma_t} M_s \cdot \frac{ds}{||\Sigma_t||}$$
Plugging in the value for $M_s$, we get
$$ M_{avg} = C \cdot \left(  \lim_{t \to \infty} \frac{1}{||\Sigma_t||} \cdot \int_{s \in \Sigma_t} \frac{1}{N_s} \, ds \right) $$
By definition of $\Sigma_t$, for all $t$ we have
$$ \int_{s \in \Sigma_t} \frac{1}{N_s} \,ds < \infty $$
Let $( \Omega, A, \mu)$ be a measure space, such that $\mu(\Omega) = 1$. If $g$ is a real-valued function that is $\mu$-integrable, and if $\varphi$ is a convex function on the real axis, then Jensen's  inequality states that \cite{rudin}
$$\varphi\left(\int_\Omega g\, d\mu\right) \le \int_\Omega \varphi \circ g\, d\mu $$
where $\varphi \circ g$ is a composition of $\varphi(.)$ and $g(.)$. Now, taking $g(s) = N_s$ , $\varphi(x)=\frac{1}{x}$ and $\Omega = \Sigma_t$, we get 
$$ \int_{s \in \Sigma_t} \frac{1}{N_s} \, \frac{ds}{||\Sigma_t||}  \geq  \frac{1}{\int_{s \in \Sigma_t} N_s \, \frac{ds}{|\Sigma_t|}}$$
Hence we get a lower bound for the long term average service rate as
\begin{eqnarray} \label{eq:jn}
 M_{avg} \geq \frac{C \cdot \displaystyle \lim_{t \to \infty} ||\Sigma_t||}{\displaystyle \lim_{t \to \infty} \int_{s \in \Sigma_t} N_s \, ds}
\end{eqnarray}
where 
$$ ||\Sigma_t|| = \int^t_0 \boldsymbol 1_{\{N_s \neq 0\}} \, ds$$
It is easy to see that
$$ \int_{s \in \Sigma_t} N_s \, ds =  \int^t_0 N_s \, ds $$
Substituting the two equations above in Equation (\ref{eq:jn}) and multiplying the numerator and denominator by $\frac{1}{t}$, we get
\begin{eqnarray} \label{eq:mavg}
M_{avg} \geq C \cdot \left(  \frac{\displaystyle \lim_{t \to \infty} \frac{1}{t} \int^t_0 \boldsymbol 1_{\{N_s \neq 0\}} \, ds}{\displaystyle \lim_{t \to \infty} \frac{1}{t} \int^t_0 N_s \, ds}  \right) 
\end{eqnarray}
Now $\{N_s, s \geq 0\}$ is a finite dimensional irreducible continuous time Markov chain, hence it is positive recurrent and ergodic. Thus there exists a stationary distribution $\boldsymbol\pi$. From the property of stationary distribution, we have
\begin{eqnarray} \label{eq:pi}
 \pi(i) = \lim_{t \to \infty} \frac{1}{t} \int^t_0 \boldsymbol 1_{\{ N_s = i \}} ds
\end{eqnarray}
Now, $N_s$ can be written alternatively as
$$N_s = \sum^n_{i=0} i \cdot \boldsymbol 1_{\{N_s = i\}} $$

\noindent
Now, 
$$\lim_{t \to \infty} \frac{1}{t} \int^t_0 N_s  ds  = \lim_{t \to \infty} \frac{1}{t} \int^t_0 \left( \sum^n_{i=0} i \cdot \boldsymbol 1_{\{N_s = i\}} \right) ds$$
Since the summation is over finite terms, by inter-changing the order of operations, we get
$$\lim_{t \to \infty} \frac{1}{t} \int^t_0 N_s  ds  = \sum^n_{i=0} i \cdot \left( \lim_{t \to \infty} \frac{1}{t} \int^t_0 \boldsymbol 1_{\{N_s = i\}} ds \right)$$
Substituting from Equation (\ref{eq:pi}),
$$\lim_{t \to \infty} \frac{1}{t} \int^t_0 N_s  ds  = \sum^n_{i=0} i \cdot \pi(i)$$

\noindent
Let $p_0$ be the steady state probability of a queue being empty. From queuing theory, it is known that
$$p_0 = 1 - \frac{\lambda}{M_{avg}}$$ 
Since we have decoupled the queues and thus assumed them to be independent, we have
$$\pi(i) = P\{i \textrm{ queues are non-empty} \}$$
$$\pi(i) = \binom{n}{i} (1-p_0)^i (p_0)^{n-i}$$
Through algebraic simplification, we get
\begin{eqnarray} \label{eq:ens}
\lim_{t \to \infty} \frac{1}{t} \int^t_0 N_s  ds = \sum^n_{i=0} i \cdot \pi(i) = n(1-p_0)
\end{eqnarray}
Again From Equation (\ref{eq:pi}), we have
$$ \lim_{t \to \infty} \frac{1}{t} \int^t_0 \boldsymbol 1_{\{N_s \neq 0\}} \, ds  = \sum^n_{i=1} \pi(i) = 1 - \pi(0)$$
Plugging in the value of $p_0$, we obtain
\begin{eqnarray} \label{eq:pi0}
\lim_{t \to \infty} \frac{1}{t} \int^t_0 \boldsymbol 1_{\{N_s \neq 0\}}  ds  = 1 - {p_0}^n
\end{eqnarray}
Substituting Equations (\ref{eq:ens}) and (\ref{eq:pi0}) in Equation (\ref{eq:mavg}), 
$$M_{avg} \geq C \cdot \frac{(1-{p_0}^n)}{n(1-p_0)}$$
Substituting the value of $p_0$ in the above equation and simplifying, we get
\begin{eqnarray} \label{eq:hpp}
M_{avg} \geq \frac{\lambda}{1 - \sqrt[n]{1-\frac{n\lambda}{C}} } 
\end{eqnarray}
Queuing theory gives us the expression for average delay as,
$$d_{avg} = \frac{1}{M_{avg}-\lambda} $$
After substitution and simplification, we obtain
$$d_{avg} \leq \frac{1}{\lambda} \cdot \left(\frac{1}{\sqrt[n]{1-\frac{n\lambda}{C}}}  - 1 \right) $$

\subsection{Modelling average delay for non-homogeneous Poisson packet generation}
In this section, we briefly show how the above result can be extended to the case when the packet generation rates are non-homogeneous Poison process. We re-write Equation \ref{eq:hpp} as
\begin{eqnarray} \label{eq:hpp1}
1 - \frac{n\lambda}{C}  \leq \left(1 - \frac{\lambda}{M_{avg}} \right)^n 
\end{eqnarray} 
We observe that in Equation \eqref{eq:hpp1}, we are ordering the probability of the the system with the decoupled queues begin empty to the probability that that the system begin empty if we aggregate all the traffic and assume it to be served at rate $C$. Let the node $i \in \{1,2,3,...,n \}$ generate packets as Poisson process with rate $\lambda_i$. By using the above argument, we extend Equation \eqref{eq:hpp1} for non-homogeneous Poisson packet generation as
\begin{eqnarray} \label{eq:npp}
1 - \frac{\displaystyle \sum^{n}_{i=1} \lambda_i}{C}  \leq \prod^n_{i=1} \left(1 - \frac{\lambda_i}{M_{avg}} \right) 
\end{eqnarray}
Equation \ref{eq:npp}, can be solved for $M_{avg}$, using fixed point method under equality. The upper bound for the mean delay for node $i$, i.e $d^{(i)}_{avg}$, can be then computed as
$$d^{(i)}_{avg} \leq \frac{1}{\lambda_i - M_{avg}}$$

\section{\large{Simulation}}
\label{sim}
In this section, we show the plots which compare the theoretical against the simulated values of mean delay. The simulation was done using Qualnet 4.5, which is a discrete event simulation system. In order to obtain accurate estimates of the mean delay, the simulation was run long enough so that the average delay at the root was within $1 \mu s$ interval. This process was continued for $30$ simulation runs to obtain a confidence interval for the mean delay with $95\%$ confidence. For theoretical computation of delay, $C$ is taken as $72.8\, pkts/s$ from Figure \ref{fig:thruput}.

\begin{figure}[!h]
\centering
\includegraphics[scale=0.22,angle=-90]{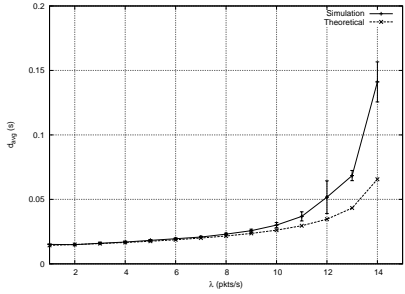}
\caption{Variation of $d_{avg}$ with $\lambda$ when 5 nodes are sending data to \textit{ROOT}. Packet size is $1500\,Bytes$ and the transmission rate is $1\,Mbps$ }
\label{fig:plot5}
\end{figure}

\newpage
In Figure \ref{fig:plot5}, we plot the variation in $d_{avg}$ against arrival rate $\lambda$. We observe that for low values of $n \lambda$, the simulated mean delay and theoretical mean delay are close to each other. We find that the following combinations of $n$ and $\lambda$ ensure that the error between the theoretical and simulated delay is within $10\%$.

\begin{center}
\begin{tabular}{|c|c|c|}
\hline $n$ & $\lambda\,(pkts/s)$ & $\frac{n\lambda}{C}$\\  
\hline 3 &  17.0 & 0.70\\ 
\hline 4 &  13.0 & 0.71\\ 
\hline 5 &  13.0 & 0.69\\
\hline 6 &  6.0 & 0.49\\
\hline 7 &  4.0 & 0.39\\
\hline 8 &  3.0 & 0.33\\
\hline 9 &  3.0 & 0.37\\
\hline 10 & 3.0  & 0.41\\
\hline
\end{tabular} 
\end{center}

\begin{figure}[!h]
\centering
\includegraphics[scale=0.22,angle=-90]{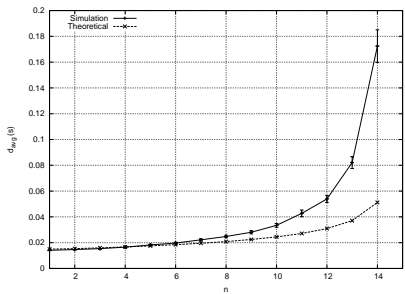}
\caption {Variation of $d_{avg}$ with $n$ when $\lambda = 5$ pkts/sec. Packet size is $1500\,Bytes$ and the transmission rate is $1\,Mbps$ }
\label{fig:rate5}
\end{figure}

Our analysis is based on the fundamental assumption that the queues evolve independently at low load, the chances of several queues being non-empty at the same time is low and intuitively the independence assumption applies. But as load increases, the interactions between the queues appear and our assumptions ceases to remain valid. This accounts for the mismatch between simulation and analysis that is observed in Figure \ref{fig:rate5} for higher values of $\lambda$.

In Figure \ref{fig:rate5}, we plot the variation in $d_{avg}$ against the number of contending nodes $n$ for a given arrival rate. We observe that the theoretical and simulated means are matching for low to moderate load regime. 

Again from the plots, we observe that the theoretical values deviates significantly from the simulation results under heavy load. With increasing load, the coupling between the queues may become stronger, thus invalidating the assumption of independent queues in our model. 

\section{\large{Conclusion and Future Work}}
In this paper, we have proposed a novel and simplified  analysis for delay modelling in single-hop wireless mesh networks. In our analysis, we have represented the system as decoupled independent M/M/1 queues. We have derive closed form expression for mean delay, when the packet generation is a homogeneous Poisson process. We have also extended our results to characterize the mean delay for the case of non-homogeneous Poisson packet generation.

Simulations indicate that even though our analysis makes apparently drastic assumptions, the results match simulations very well when the load on the network is small to moderate. For example, in the light to moderate load regime, the error between analysis and simulation is not more than 10\%. Thus for low to moderate loads, we have a simple analytical formula that captures mean delay with reasonable accuracy. Nevertheless, we consider a model in this paper as a first step in analyzing the problem.

Our on-going work is concerned with refining the model; we are seeking a tractable approach that does not need to assume that each queue evolves independently. Subsequently, we would like to accommodate non-homogeneous and non-Poisson arrivals; this may require modelling the queues as G/G/1 queues. We would also like to extend this to multi-hop wireless mesh networks.

\bibliography{delay_modelling_mm1_final}

\end{document}